\documentclass[journal,transmag]{IEEEtran}
\usepackage{microtype}
\usepackage{graphicx}
\usepackage{subfigure}
\usepackage{booktabs} 
\usepackage{multirow}
\usepackage{amsmath}
\usepackage{amssymb}
\usepackage{amsthm}
\usepackage{setspace}
\usepackage{pgfplots}
\usetikzlibrary{pgfplots.groupplots}
\usepackage{mathtools}\mathtoolsset{showonlyrefs}
\usepackage{algorithm}
\usepackage{algorithmic}
\usepackage{hyperref}

\usepackage{lipsum} 
\ifCLASSINFOpdf
\else
\fi
\begin{document}

\newcommand{\myhl}[1]{\textcolor{black}{#1}}
\newcommand{\myhltwo}[1]{\textcolor{black}{#1}}
%
\title{Gated Recurrent Context: Softmax-free \myhltwo{A}ttention \\ 
for Online Encoder-Decoder Speech Recognition}


\author{\IEEEauthorblockN{Hyeonseung Lee, 
Woo Hyun Kang, 
Sung Jun Cheon, 
Hyeongju Kim, 
\\and 
Nam Soo Kim,
 ~\IEEEmembership{Senior Member,~IEEE}}
\thanks{

The authors are with the Institute of New Media and Communications, Department of Electrical and Computer Engineering, Seoul National University, Seoul, Republic of Korea (e-mail: hslee@hi.snu.ac.kr; whkang@hi.snu.ac.kr; sjcheon@hi.snu.ac.kr; hjkim@hi.snu.ac.kr; nkim@snu.ac.kr). 


© 2021 IEEE.  Personal use of this material is permitted.  Permission from IEEE must be obtained for all other uses, in any current or future media, including reprinting/republishing this material for advertising or promotional purposes, creating new collective works, for resale or redistribution to servers or lists, or reuse of any copyrighted component of this work in other works.

Digital Object Identifier 10.1109/TASLP.2021.3049344
}}

\markboth{
}%
{Shell \MakeLowercase{\textit{et al.}}: Bare Demo of IEEEtran.cls for IEEE Transactions on Magnetics Journals}
%



\maketitle
\begin{abstract}
Recently, attention-based encoder-decoder (AED) models have shown state-of-the-art performance in automatic speech recognition (ASR). As the original AED models with global attentions are not capable of online inference, various online attention schemes have been developed to reduce ASR latency for better user experience. However, a common limitation of the conventional softmax-based 
online attention approaches is that they introduce 
an additional hyperparameter related to the length of the attention window, requiring multiple trials of model training for tuning the hyperparameter. In order to deal with this problem, we propose a novel softmax-free 
attention method and its modified formulation for online attention, which does not need any additional hyperparameter at the training phase. Through a number of ASR experiments, we demonstrate the tradeoff between the latency and performance of the proposed online attention technique can be controlled by merely adjusting a threshold at the test phase. Furthermore, the proposed methods showed competitive performance to the conventional 
global and online 
attentions in terms of word-error-rates (WERs).
\end{abstract}



\IEEEdisplaynontitleabstractindextext

%
\IEEEpeerreviewmaketitle

\section{Introduction}
%
%
%
%
\label{sec:introduction}

\IEEEPARstart{I}{n} the last few years, the performance of deep learning-based end-to-end automatic speech recognition (ASR) systems has improved significantly through numerous studies mostly on the architecture designs and training schemes of neural networks (NNs).
Among many end-to-end ASR systems, attention-based encoder-decoder (AED) models~\cite{LAS, Bahdanau} have shown better performance than the others, such as the connectionist temporal classification (CTC)~\cite{CTC} and recurrent neural network transducer (RNN-T)~\cite{RNN-T}, and even outperformed the conventional DNN-hidden Markov model (HMM) hybrid systems 
in case a large training set of transcribed speech is available~\cite{SOTA-LAS}. 
Such successful results of AED models come from the tightly integrated language modeling capability of the label-synchronous decoder \myhl{(i.e., the decoder network operates once per output text token in an autoregressive manner)}, supported by the attention mechanism that provides proper acoustic information at each step~\cite{garg2019improved}. 

A major drawback of the conventional AED models is that they cannot infer the ASR output in an online fashion, 
which 
degrades the user experience due to the large latency~\cite{sainath2019two}. This problem is mainly caused by the following aspects of the AED models. 
Firstly, the encoders of most high-performance AED models make use of layers with global receptive fields, such as bidirectional long short-term memory (BiLSTM) or self-attention layer. 
More importantly, a conventional global attention mechanism (e.g., Bahdanau attention) considers the entire utterance to obtain the attention context vector at every step.
The former issue can be solved by replacing the global-receptive encoder with an online encoder%
, where an encoded representation 
for a particular frame depends on only a limited number of future frames. The online encoder can be built straightforwardly by employing layers with finite future receptive field such as latency-controlled BiLSTM (LC-BiLSTM)~\cite{zhang2016highway}, \myhl{temporal convolution layers,} and masked self-attention layers. 
However, reformulating the global attention methods for an online purpose is still a challenging problem.


Conventional techniques for online attention
are usually two-step approaches 
where the window (i.e., chunk) for the current attention is determined first at each decoder step, then the attention weights are calculated using the softmax function  
defined over the window.
\myhl{Existing online attentions mainly 
differ in 
how 
they determine the window.}
Neural transducers~\cite{jaitly2015neural, sainath2018improving} divide an encoded sequence into multiple chunks with a fixed length, and the attention-decoder produces an output sequence for each input chunk.
In the windowed attention 
techniques~\cite{hou2017gaussian, tjandra2017local, merboldt2019analysis}, the 
position of each fixed-size window is decided by a position prediction algorithm. The window position is monotonically increasing in time, and some approaches employ a trainable position prediction model with a fixed Gaussian window.
In MoChA-based approaches~\cite{chiu2018monotonic,miao2019online, tsunoo2019towards}, a fixed-size chunk is obtained using 
a monotonic endpoint prediction model, which is jointly trained 
considering all possible 
chunk endpoints. 

A common limitation of the aforementioned approaches is 
that the 
fixed-length of the window needs to be tuned according to the training data.
Merely choosing a large 
window of a constant size causes a large latency while setting the window size too small results in degraded performance. 
Therefore multiple trials of the model training are required to find a proper value of the window length, 
consuming 
excessive computational resources. 
Moreover, the trained model 
does not guarantee to perform well on an unseen test set, since the window size is fixed for all datasets. 

Although a few variants of MoChA utilize an adaptive window length to remove the need for tuning the window size, such variants induce other problems.
MAtChA~\cite{chiu2018monotonic} regards the previous endpoint as the beginning of the current chunk. Occasionally, the window can be too short to contain enough speech content when two consecutive endpoints are too close, which may degrade the performance. 
AMoChA~\cite{fan2018online} employs an auxiliary model that predicts the chunk size but also introduces 
an additional loss term for the prediction model. As the coefficient for the new loss needs tuning, AMoChA still requires 
repeated training sessions. 
Besides, several recent approaches~\cite{moritz2019triggered,dong2019cif} utilize strictly monotonic windows.
But these methods have a limitation in that the decoder state is not used for determining the window, which means such algorithms might not fully exploit the advantage of AED models, i.e., the inherent capability of autoregressive language modeling.

The 
aforementioned inefficiency in training the conventional online attentions 
is essentially caused by the fact that the softmax function 
needs a predetermined attention window to obtain the attention weights
, which 
results in repetitive tuning process of the window-related hyperparameter.
\myhl{Although several recent studies~\cite{tsai2019transformer, katharopoulos2020transformers} investigate softmax-free formulation of attention, they focuses on reducing computations by replacing the softmax with other kernels and do not suggest a solution for online encoder-decoder attention.}
To overcome this limitation
, we propose a novel softmax-free global attention method called gated recurrent context (GRC)
, inspired by the 
gate-based update 
in gated recurrent unit (GRU)~\cite{cho2014learning}. 
Whereas conventional attentions are based on a kernel smoother (e.g. softmax function)~\cite{wasserman2006all,tsai2019transformer}, GRC obtains an attention context vector by recursively aggregating the encoded vectors in a time-synchronous manner, using 
update gates. 
GRC can be 
reformulated 
for the purpose of online attention, 
which we refer to decreasing GRC (DecGRC), where the update gates are constrained to be decreasing over time. DecGRC is \myhl{window-free} and capable of deciding the attention-endpoint by thresholding the update gate values at the inference phase.
DecGRC as well as GRC introduces no hyperparameter to be tuned at the training phase.

The main contributions of this paper can be summarized as follows:

\begin{itemize}
\item We propose a novel softmax-free attention method called \textbf{Gated Recurrent Context (GRC)}
\myhl{, which obtains an attention context vector using a time-synchronous recursive updating rule rather than a kernel smoother-based formulation.}
%
\item We present a 
\myhl{window-free} online attention method, \textbf{Decreasing GRC (DecGRC)}, a constrained 
variant of GRC
. DecGRC 
does not need 
any new hyperparameter to be tuned at the training phase. 
At test time, the tradeoff between performance and latency can be adjusted using a  
simple thresholding technique.
\item We experimentally 
show that GRC and DecGRC 
perform competitive to the conventional global and online 
attention methods on the LibriSpeech test set.
\end{itemize}

The remainder of this paper is organized as follows. In Section~\ref{sec:backgrounds}, the general framework of attention-based encoder-decoder ASR is formally described, followed by conventional online attention methods and their common limitation.
Section~\ref{sec:proposed} proposes formulations of both GRC and DecGRC and the algorithm for online inference. The experimental results with 
various attention methods are 
given in Section~\ref{sec:experiments}. 
Conclusions are presented in Section~\ref{sec:conclusion}.
 


\begin{figure*}[t]
\centerline{\includegraphics[width=19cm]{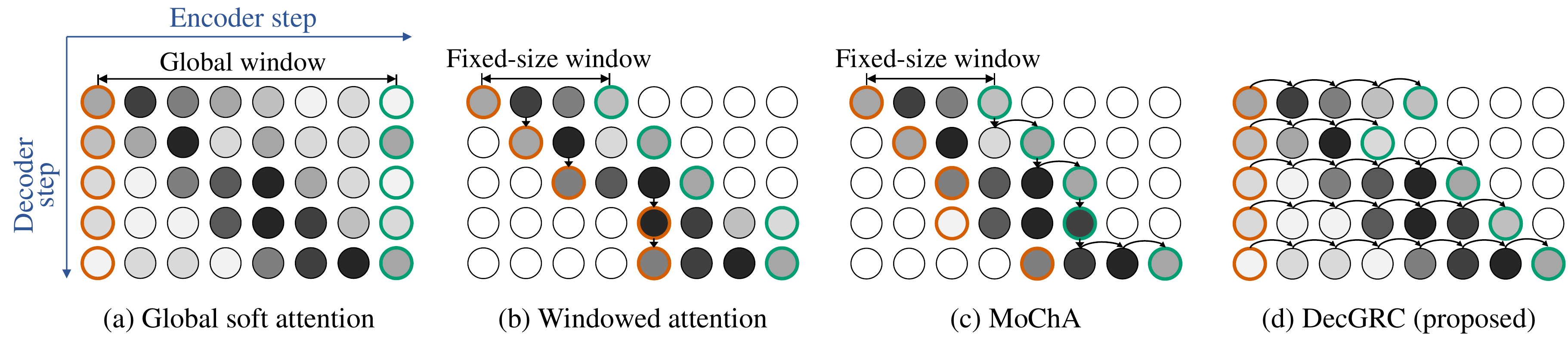}}
\vskip 0.1in
\caption{\myhl{Pictorial descriptions of various attention methods. For online attention methods, the endpoint and the start-point (if it exists) are respectively marked with cyan and orange bold outline, at each decoder step.} 
\myhltwo{Windowed attention and MoChA, two widely-used online attention methods, decide either the start-point or endpoint for each decoder step, and then calculate attention weights within a fixed-size window. 
Unlike these conventional methods, DecGRC does not utilize a fixed-size window and scans from the beginning of the utterance to find the endpoint for each decoder step. The endpoint decision algorithm of DecGRC is independent of the former endpoints. Thus the endpoint may not be monotonically increasing over time-steps, as depicted in (d). The detailed DecGRC inference algorithm is described in Alg.~\ref{alg:example}. }} 
\label{attention_description}
\end{figure*}

\section{Backgrounds}
\label{sec:backgrounds}
\subsection{Attention-based Encoder-Decoder for ASR}
\label{AED_based_ASR}
An attention-based encoder-decoder model consists of two sub-modules $\mathrm{Encoder}(\cdot)$ and $\mathrm{AttentionDecoder}(\cdot)$, and it predicts the posterior probability of the output transcription given the input speech features as follows:
\begin{equation}
\mathbf{h} = \mathrm{Encoder}(\mathbf{x}),
\end{equation}
\begin{equation}
P(\mathbf{y}|\mathbf{x}) = \mathrm{AttentionDecoder}(\mathbf{h},\mathbf{y})
\end{equation}
where $\mathbf{x}=[x_1, x_2, ..., x_{T_{in}} ]$ and $\mathbf{h}=[h_1, h_2, ..., h_{T} ]$ are sequences of
input speech features and encoded vectors respectively, and $\mathbf{y}=[y_1, y_2, ..., y_U ]$ is a sequence of output text units. Either the start or end of the text 
is considered as one of the text units.

In general, 
$\mathrm{Encoder}(\cdot)$ reduces 
its output length $T$ to be smaller than the input length $T_{in}$, cutting down the memory and computational footprint. 
A global $\mathrm{Encoder}(\cdot)$ is 
implemented with NN layers having 
powerful sequence modeling capacity, e.g., BiLSTM or self-attention layers with 
subsampling layers. On the other hand, an online $\mathrm{Encoder}(\cdot)$ must only consist of layers with finite future receptive field.

$\mathrm{AttentionDecoder}(\cdot)$ operates at each output step recursively, emitting an estimated posterior probability over all possible text units given 
the outputs produced at the previous step.
This procedure can be summarized as follows:
\begin{equation}
s_u =\mathrm{RecurrentState}(s_{u-1}, y_{u-1}, c_{u-1}),
\end{equation}
\begin{equation}
\label{AttentionContext}
c_u = \mathrm{AttentionContext}(s_u, \mathbf{h}),
\end{equation}
\begin{equation}
P(y_u | \mathbf{y}_{<u}, \mathbf{x}) = \mathrm{ReadOut}(s_u, y_{u-1}, c_u)
\end{equation}
where $c_u$ denotes the $u$-th attention context vector and $s_u$ is the $u$-th decoder state. 
$\mathrm{RecurrentState}(\cdot)$ consists of unidirectional layers, e.g., unidirectional LSTM and masked self-attention layers. 
$\mathrm{ReadOut}(\cdot)$ usually contains a small NN followed by a softmax activation function. 

The most popular choice for $\mathrm{AttentionContext}(\cdot)$ is the global soft attention (GSA)~\cite{Bahdanau, luong2015effective} that includes the softmax function given as follows:
\begin{equation}
\label{eq:GSA_context}
c_u = \sum_{t=1}^{T}{\alpha_{u,t}} h_t ,
\end{equation}
\begin{equation}
\label{GSA_weights}
\alpha_{u,t} = \frac{\exp(e_{u,t})} {\sum_{j=1}^{T}{\exp(e_{u,j})}},
\end{equation}
\begin{equation}
\label{GSA_score}
e_{u,t} =  \mathrm{Score}(s_u, h_t, \alpha_{<u,t})
\end{equation}
in which $\alpha_{u,t}$ is an attention weight on the $t$-th encoded vector $h_t$ at the $u$-th decoder step, and $e_{u,t}$ is a score indicating the relevance of $h_t$ to the $u$-th decoder state. \myhl{Common choices for the $\mathrm{Score}(\cdot)$ function are additive scores~\cite{Bahdanau,zeyer2018improved} and dot-product scores~\cite{luong2015effective,vaswani2017attention}. Additive scores often utilize additional information $\alpha_{<u,t}$ to decide the current attention weights based on the past attention locations. In this paper, an additive score with attention weight feedback~\cite{zeyer2018improved} is employed for all the experiments at Sec.~\ref{sec:experiments}:
\begin{equation}
\label{eq:score_exp1}
e_{u,t} = v^T \mathrm{tanh}(W[s_u; h_t; \beta_{u,t}] + \eta),
\end{equation} 
\begin{equation}
\label{eq:score_exp2}
\beta_{u,t} = \sigma(v_\beta^{T} h_t) \cdot \sum_{k=1}^{u-1} \alpha_{k,t}
\end{equation}
where the notation $[\,\cdot\,;\,\cdot\,]$ means concatenation of vectors, $v$ and $v_\beta$ are trainable vectors, $W$ and $\eta$ are a trainable weight and a trainable bias, and $\beta_{u,t}$ is an attention weight feedback.
}

The whole system is trained to maximize the log posterior probability on a training dataset $\mathbf{D}=\{(\mathbf{x}^{(n)},\mathbf{y}^{(n)}) \}_{n=1}^{N} $, 
\begin{equation}
\mathrm{max}_{\theta} \,\, \mathrm{E}_{ (\mathbf{x},\mathbf{y}) \sim \mathbf{D}} \Big[ 
\sum_{u=1}^{{|\mathbf{y}|}}{\log}P(y_u| \mathbf{y}_{<u}, \mathbf{x} ;\theta) \Big]
\end{equation}
where $\theta$ denotes the set of 
all trainable parameters, and $|\mathbf{y}|$ 
is the 
text sequence length of the sampled data.
Inference can be performed by searching the most likely text sequence:
\begin{equation}
\mathbf{\hat{y}} = \mathrm{argmax}_{\theta} \,\, {\log}P(\mathbf{y} | \mathbf{x} ;\theta).
\end{equation}

\subsection{Online Attention}

\label{sec:online_attention}
To achieve online attention, 
the context vector $c_u$ in Eq.~\eqref{eq:GSA_context} must have local dependency on the encoded vectors $\mathbf{h}$.
Windowed attention and MoChA are 
widely-used online attention methods that show high performance 
for which only the $\mathrm{AttentionContext}(\cdot)$ function in Eq.~\eqref{AttentionContext} is modified in 
the general framework. 
\myhl{Pictorial descriptions of all the online attention methods in this paper are provided in Fig.~\ref{attention_description}.}

\subsubsection{Windowed attention}
Among various formulations of windowed attention, a simple heuristic using $\mathrm{argmax}$ for window boundary prediction~\cite{merboldt2019analysis} has shown the best performance. This 
method can be described as follows:
\begin{equation}
\label{Windowed_att_position}
p_1 = 0, \quad p_u = \mathrm{argmax}_{t}(\bold{\alpha}_{u-1, 1\leq t\leq T}),
\end{equation}
\begin{equation}
\label{Windowed_att_context}
c_u = \sum_{t = p_u}^{p_u+w-1}{\alpha_{u,t}} h_t ,
\end{equation}
\begin{equation}
\label{Windowed_att_weights}
\alpha_{u,t} =  \frac{\exp(e_{u,t})} {\sum_{j=p_u}^{p_u+w-1}{\exp(e_{u,j})}}
\end{equation}
where $p_u$ is the start point of the attention window at the $u$-th step, and $w$ is the window size.
The windowed attention is online, as the attention context $c_u$ derived through Eqs.~\eqref{Windowed_att_position}-
\eqref{Windowed_att_weights} does not depend on the 
entire encoded vector sequence $\mathbf{h}$. 
The tradeoff between performance and latency of windowed attention relies on the window length 
$w$. 

\subsubsection{MoChA}
In 
MoChA~\cite{chiu2018monotonic}, 
an attention window endpoint is first decided, 
followed by attention weights calculation 
within a fixed-
size window as follows:
\begin{equation}
\label{MoChA_infer_context}
c_{u} = \sum_{t=\tau_u-w+1}^{\tau_u}\beta_{u,t} h_t,
\end{equation}
\begin{equation}
\label{MoChA_infer_beta}
\beta_{u,t} = \frac{\exp(e_{u,t})} {\sum_{j=\tau_u-w+1}^{\tau_u}{\exp(e_{u,j})}},
\end{equation}
\begin{equation}
\label{MoChA_infer_tau}
\tau_{u} = \mathrm{MonotonicEndpoint}(\tilde{e}_{u,\geq\tau_{u-1}}) ,
\end{equation}
\begin{equation}
\label{Monotonic_infer_chunkenergy}
\tilde{e}_{u,t} =  \mathrm{MonotonicScore}(s_u, h_t, \beta_{<u,t}) + b
\end{equation}
where $\mathrm{MonotonicScore(\cdot)}$ is a similarity function, $b$ is a trainable bias parameter, $\tilde{e}_{u,t}$ is the monotonic score, and $\mathrm{MonotonicEndpoint(\cdot)}$ is an window end-decision algorithm based on thresholding, and $\beta_{u,t}$ is an attention weight within the window. Note that Eqs.~\eqref{MoChA_infer_context}-
\eqref{Monotonic_infer_chunkenergy} 
are substitutes for Eqs.~\eqref{eq:GSA_context}-
\eqref{GSA_weights} in GSA. The performance and latency of MoChA are also known to depend on the chunk size $w$.

Optimizing an AED model using these formulations is \myhl{impossible. The $\mathrm{MonotonicEndpoint(\cdot)}$ function makes a hard-decision for an endpoint $\tau_u$ that is not 
differentiable, which means $\tau_u$ cannot be trained with the backpropagation framework.}
To solve this problem, an expectation-based formulation is exploited for training~\cite{chiu2018monotonic}:
\begin{equation}
\label{MoChA_train_beta}
\beta_{u,t} = \sum_{k=t}^{t+w-1} \frac{ \alpha_{u,t} \exp(e_{u,k})}{  \sum_{l=k-w+1}^{k}\exp(e_{u,l}) },
\end{equation}
\begin{equation}
\label{MoChA_train_alpha}
\alpha_{u,t} = p_{u,t} \Big( (1-p_{u,t-1})\frac{\alpha_{u,t-1}}{p_{u,t-1} } + \alpha_{u-1,t} \Big),
\end{equation}
\begin{equation}
\label{MoChA_train_p}
p_{u,t} = \sigma(e_{u,t})
\end{equation}
where $p_{u,t}$ is a stopping probability at the $t$-th time step and $\alpha_{u,t}$ is an accumulated selection probability that the window endpoint is $t$. 

\subsubsection{A limitation of the conventional methods}
As mentioned in Sec.~\ref{sec:introduction}, 
the softmax function in the conventional online attentions (e.g., Eqs.~\eqref{Windowed_att_weights}-
\eqref{MoChA_infer_beta}) 
requires a predetermined attention window, 
which induces a limitation 
in training efficiency since multiple trials of training are inevitable for tuning
either the window length or the coefficient of an additional loss term.
To overcome this limitation, in the next section, we propose a novel softmax-free global attention approach and its online version which is free 
from the tuning of hyperparameters in training.

\section{Proposed methods}
\label{sec:proposed}
\subsection{Gated Recurrent Context (GRC)}
We propose a novel softmax-free global attention method called GRC, which recursively aggregates the information of the encoded sequence into an attention context vector in a time-synchronous manner. 
Specifically, the following formulas are employed in place of the Eqs.~\eqref{eq:GSA_context}-
\eqref{GSA_score}:
\begin{equation}
\label{GRC_context}
c_{u} = d_{u,T},
\end{equation}
\begin{equation}
\label{GRC_recursive}
\quad d_{u,t} = (1-z_{u,t})d_{u,t-1} + z_{u,t}h_t,
\end{equation}
\begin{equation}
\label{GRC_updategate}
z_{u,1}=1, \quad z_{u,t} = \sigma(e_{u,t}) = \frac{1}{1+\exp(e_{u,t})},
\end{equation}
\begin{equation}
\label{GRC_energy}
e_{u,t} = \mathrm{Score}(s_u, h_t, \alpha_{<u,t}) + b
\end{equation}
where $z_{u,t}$ and $d_{u,t}$ are the update gate and the intermediate attention context vector for the $t$-th time step at the $u$-th decoder step, respectively. 
GRC computes an intermediate value for the final context vector recursively in time, inspired by GRU~\cite{cho2014learning}. 
Note that Eqs.~\eqref{GRC_context}-
\eqref{GRC_energy} of GRC do not utilize the softmax function 
at all, unlike the conventional attentions. 
Nevertheless, GRC can be interpreted as a global attention method, 
since it calculates a weighted average of the encoded sequence over the whole time period, as explained in Sec.~\ref{sec:relation_to_GSA}.

\subsubsection{Relation to GSA}
\label{sec:relation_to_GSA}
\myhl{The update gates sequence $\bold{z}_{u,1:T}$ of GRC in Eq.~\eqref{GRC_updategate} and the attention weights sequence $\bold{\alpha}_{u,1:T}$ of GSA in Eq.~\eqref{GSA_weights} }
have one-to-one correspondence \myhl{(i.e., intuitively, $\bold{z}_{u,1:T}$ and $\bold{\alpha}_{u,1:T}$ are always interchangable without changing the value of attention context vector $c_{u}$)} according to 
the following theorem: 
\newtheorem{theorem}{Theorem}
\newtheorem{corollary}{Corollary}[theorem]
\begin{theorem}[GRC-GSA duality]
\label{thm1}
For arbitrary $n \in \mathbb{N}$, let $Z^n=\{x \in \mathbb{R}^{n} \,|\,x_1=1,\,\, 0 \leq x_j \leq 1 \,\, for \,\, j=2,3,\dots,n \}$ and $A^n=\{x \in \mathbb{R}^{n} \,|\,\sum_{j=1}^{\myhl{n}}x_j=1,\,\, 0 \leq x_j \leq 1 \,\, for \,\, j=1,2,\dots,n \}$. 
There exists a bijective function $\boldsymbol{\bar{\alpha}} :Z^T \rightarrow A^T$ s.t. for any $\mathbf{h}=[h_1, h_2, \dots, h_T]$ and $\mathbf{z}_{u}=[z_{u,1}, z_{u,2}, \dots, z_{u,T}]$, the following holds: 
\begin{equation}
\label{thm1_eq}
d_{u,T} = \sum_{t=1}^{T} \boldsymbol{\bar{\alpha}}(\mathbf{z}_{u})_{t}h_t
\end{equation}
where $\boldsymbol{\bar{\alpha}}(\mathbf{z}_{u})_{t}$ denotes the $t$-th element of $\boldsymbol{\bar{\alpha}}(\mathbf{z}_{u})$, and $d_{u,T}$ is 
obtained from $\mathbf{z}_{u}$ and $\mathbf{h}$ according to Eq.~\eqref{GRC_recursive}.

\begin{proof}[\textbf{Proof 
}]
Using the recursive Eq. \eqref{GRC_recursive},
\begin{equation} \label{thm1_proof_eq1}
\begin{split}
d_{u,T}  = &(1-z_{u,T}) d_{u,T-1} + z_{u,T}h_T \\
         = &(1-z_{u,T})(1-z_{u,T-1}) d_{u,T-2} \\
           &+ (1-z_{u,T})z_{u,T-1} h_{T-1} + z_{u,T}h_T \\
         = &\dots \\
         = &\sum_{t=1}^T \Big(\prod_{j=t+1}^{T} (1-z_{u,j})\Big)z_{u,t}h_t .
\end{split} 
\end{equation}
Therefore the function $\boldsymbol{\bar{\alpha}}$ that satisfies Eq. \ref{thm1_eq} is given by
\begin{equation} 
\label{GRC_GSA_dual_function}
\boldsymbol{\bar{\alpha}}(\mathbf{z}_{u})_t:= z_{u,t}\prod_{j=t+1}^{T} (1-z_{u,j})  \quad \text{for $t=1, 2, \dots, T$.}
\end{equation}
Given that $\mathbf{z}_u \in Z^T$, the output $\boldsymbol{\bar{\alpha}}(\mathbf{z}_{u})$ is an element of $A^T$ because it is trivial to show that $0 \leq \boldsymbol{\bar{\alpha}}(\mathbf{z}_{u})_t \leq 1$ \quad for \,\, $i = 1, 2, \dots, T$, and also $\sum_{t=1}^{T} \boldsymbol{\bar{\alpha}}(\mathbf{z}_{u})_t=1$ holds as follows:
\begin{equation} 
\begin{split}
\sum_{t=1}^{T} \boldsymbol{\bar{\alpha}}(\mathbf{z}_{u})_t &= \sum_{t=1}^{T} z_{u,t}\prod_{j=t+1}^{T} (1-z_{u,j}) \\
    &= \sum_{t=2}^{T} z_{u,t}\prod_{j=t+1}^{T} (1-z_{u,j}) + \prod_{j=2}^{T} (1-z_{u,j})\\
    &= \sum_{t=3}^{T} z_{u,t}\prod_{j=t+1}^{T} (1-z_{u,j}) + \prod_{j=3}^{T} (1-z_{u,j})\\
    &= \dots \\
    &= z_{u,T} + (1-z_{u,T}) = 1.
\end{split} 
\end{equation}
The $\boldsymbol{\bar{\alpha}}(\mathbf{z}_{u})$ is a bijective function since the inverse mapping of $\boldsymbol{\bar{\alpha}}$ exists as follows:
\begin{equation*} 
\begin{split}
z_{u,T}   & = \boldsymbol{\bar{\alpha}}(\mathbf{z}_{u})_T, \\
z_{u,T-1} & = 
  \begin{dcases}
    \frac{\boldsymbol{\bar{\alpha}}(\mathbf{z}_{u})_{T-1}}{\big(1-z_{u,T}\big)} = \frac{\boldsymbol{\bar{\alpha}}(\mathbf{z}_{u})_{T-1}}{1-\boldsymbol{\bar{\alpha}}(\mathbf{z}_{u})_T} & \text{if $\boldsymbol{\bar{\alpha}}(\mathbf{z}_{u})_T < 1$;} \\
    0 & \text{otherwise,} \\
  \end{dcases} \\
z_{u,T-2} & = 
  \begin{dcases}
    \frac{\boldsymbol{\bar{\alpha}}(\mathbf{z}_{u})_{T-2}}{1 - \boldsymbol{\bar{\alpha}}(\mathbf{z}_{u})_T - \boldsymbol{\bar{\alpha}}(\mathbf{z}_{u})_{T-1}} & \text{if 
    $\sum_{j=T-1}^{T}\boldsymbol{\bar{\alpha}}(\mathbf{z}_{u})_{j}$;
    } \\
    0 & \text{otherwise,} \\
  \end{dcases} \\
 &  \vdots\\
\therefore z_{u,t} & = 
  \begin{dcases}
    \frac{\boldsymbol{\bar{\alpha}}(\mathbf{z}_{u})_{t}}{1 - \sum_{j=t+1}^{T} \boldsymbol{ \bar{\alpha}}(\mathbf{z}_{u})_j} & \text{if $\sum_{t=j+1}^{T} \boldsymbol{\bar{\alpha}}(\mathbf{z}_{u})_{T} < 1$;} \\
    0 & \text{otherwise,} \\
  \end{dcases} 
\end{split} 
\end{equation*}
for $t= 1, 2, \dots, T$. It is also trivial to show that $z_{u,1} = 1$ and $0 \leq z_{u,t}\leq 1$ \quad for \,\, $i = 2, \dots, T$,  given that $\boldsymbol{\bar{\alpha}}(\mathbf{z}_{u}) \in A^T$. Therefore, $\mathbf{z}_u \in Z^T$.
\end{proof}
\end{theorem}
\vskip +0.1in


Note that $\boldsymbol{\bar{\alpha}}(\mathbf{z}_u)_t$ in Eq. 
\eqref{GRC_GSA_dual_function}
corresponds to the attention weight $\alpha_{u,t}$ in Eq.~\eqref{GSA_weights} of GSA. By Thm.~\ref{thm1}, the attention context vector $c_u$ of GRC is capable of expressing all possible weighted averages of the encoded representations over time, as in the GSA. Thus the range of $c_u$ in GRC or GSA is the same. Nonetheless, we empirically showed that GRC performs comparable to or even better than GSA, and the experimental results are given in Sec.~\ref{sec:experiments}. 
 
\subsubsection{Relation to sMoChA}
\label{relation_sMoChA}
The sMoChA~\cite{miao2019online} is a variant of MoChA where Eq.~\eqref{MoChA_train_alpha} is replaced by the following formula:
\begin{equation}
\label{sMoChA}
\alpha_{u,t} = p_{u,t} \prod_{j=1}^{t-1} (1-p_{u,j})
\end{equation}
which enables the optimization process 
to be more stabilized. 
Eq.~\eqref{sMoChA} is almost 
similar to the function $\boldsymbol{\bar{\alpha}}$ in Eq.~\eqref{GRC_GSA_dual_function}
, and implies evidence on the stability of GRC training. Despite this fact, 
sMoChA is an 
algorithm 
independent of GRC, as Eq.~\eqref{sMoChA} is merely used as the selection probability component in the whole training formulas and not even used for inference.

\subsection{Decreasing GRC (DecGRC)}
\myhl{By Thm.~\ref{thm1}, the final context vector $d_{u,T}$ of GRC in Eq.~\eqref{GRC_context} can be interpreted as a weighted average of encoded vectors $\bold{h}_{u,1:T}$. Thus GRC can be regarded as a kind of global attention method. Furthermore, not only the final context vector $d_{u,T}$ of GRC but also 
an 
intermediate context $d_{u,t}$ 
is a weighted average of the encoded vectors $\bold{h}_{u,1:T}$} 
according to the following corollary: 
\begin{corollary}[Weighted average]
\label{coroll1_thm1}
Let $Z^n$ and $A^n$ be the sets defined in Thm.~\ref{thm1}. For any $\tau \in \{1, 2, \dots, T\}$, $\mathbf{z}_u \in Z^\tau$, and $\mathbf{h}=[h_1, h_2, \dots, h_T]$ , there exists 
a function $\boldsymbol{\bar{a}} :Z^\tau \rightarrow A^T$ that satisfies the following equation: 
\begin{equation}
\label{coroll1_thm1_eq}
d_{u,\tau} = \sum_{t=1}^{T} \boldsymbol{\bar{a}}(\mathbf{z}_u)_{t}h_{t}
\end{equation}
where $\boldsymbol{\bar{a}}(\mathbf{z}_u)_{t}$ denotes the $t$-th element of $\boldsymbol{\bar{a}}(\mathbf{z}_u)$, and $d_{u,\tau}$ is obtained 
from $\mathbf{z}_{u}$ and $\mathbf{h}$ according to Eq.~\eqref{GRC_recursive}.
\begin{proof}[\textbf{Proof 
}]
By substituting every $T$ in the proof of Thm. \ref{thm1} with $\tau$, there exists a bijective function $\boldsymbol{\bar{\alpha}}: Z^\tau \rightarrow A^\tau$ given by 
\begin{equation} 
\boldsymbol{\bar{\alpha}}(\mathbf{z}_{u})_t:= z_{u,t}\prod_{j=t+1}^{\tau} (1-z_{u,j})  \quad \text{for $t=1, 2, \dots, \tau$,}
\end{equation}
such that 
\begin{equation*}
d_{u,\tau} = \sum_{t=1}^\tau \boldsymbol{\bar{\alpha}}(\mathbf{z}_{u})_t h_t.
\end{equation*}

It is trivial to show that the following function $\boldsymbol{\bar{a}}: Z^\tau \rightarrow A^T$ satisfies the equation in Coroll. \ref{coroll1_thm1_eq}:
\begin{equation*}
\boldsymbol{\bar{a}}(\mathbf{z}_u)_t =
  \begin{dcases}
     \boldsymbol{\bar{\alpha}}(\mathbf{z}_u)_t & \text{if $t \leq \tau$;} \\
    0 & \text{otherwise,} \\
  \end{dcases} \quad \text{for $t= 1, 2, \dots, T$.} \\
\end{equation*}

\end{proof}
\end{corollary}

Global attention methods including GSA and GRC cannot compute the attention weights without the entire sequence of the encoded vectors $\mathbf{h}$. 
However, considering that the attention techniques are methods for calculating the weighted average of the encoded vectors, 
Coroll.~\ref{coroll1_thm1} enables us to treat 
an intermediate context $d_{u,t}$ as a substitute for the attention context vector $c_u$ in Eq.~\eqref{GRC_context} of GRC even when 
the whole encoded sequence is not provided.

\definecolor{gray}{rgb}{0.4,0.4,0.4}
\begin{algorithm}[tb]
   \caption{Online inference using DecGRC.}
   \label{alg:example}
\begin{algorithmic}
   \STATE {\bfseries Input:} encoded vectors $\mathbf{h}$ of length $T$, threshold $\nu$
   \STATE {\bfseries State:} $s_0=\vec{0}$, $u=1$, $y_0=\mathrm{StartOfSequence}$
   \WHILE{$y_{u-1} != \mathrm{EndOfSequence}$} 
   \STATE $d_{u} = h_1$
   \FOR{$t=2$ {\bfseries to} $T$}
   \STATE $e_{u,t} = \mathrm{Score}(s_u, h_t,\alpha_{u,t})+b$
   \STATE $z_{u,t} = 1 / (1+\sum_{j=1}^{t}\exp(e_{u,j}))$
   \STATE $d_{u} = (1-z_{u,t})  d_{u} + z_{u,t}  h_t$
   \IF{$z_{u,t} < \nu $}
   \STATE \textbf{break}
   \ENDIF
   \ENDFOR
   \STATE $s_u =\mathrm{RecurrentState}(s_{u-1}, y_{u-1}, d_{u})$
   \STATE $\tilde{P}(y_u | \mathbf{y}_{<u}, \mathbf{h}) = \mathrm{ReadOut}(s_u, y_{u-1}, c_u)$ \textcolor{gray}{// softmax} 
   \STATE $y_u = \mathrm{Decide}\big(\tilde{P}(y_u | \mathbf{y}_{<u}, \mathbf{h})\big)$ \textcolor{gray}{// choose a text unit in the vocabulary. (e.g. argmax for greedy search)} 
   \STATE $u=u+1$
   \ENDWHILE
\end{algorithmic}
\end{algorithm}

Inspired by this, we further propose a novel online attention algorithm, namely DecGRC. 
DecGRC is a modified version of GRC, replacing Eq.~\eqref{GRC_updategate} with 
\begin{equation}\label{DecGRC}
z_{u,1}=1, \quad z_{u,t} = \frac{1}{1+\sum_{j=1}^{t}\exp(e_{u,j})}.
\end{equation}
Note that the update gate is constrained to be monotonically decreasing over time.
\myhl{
At the training phase, DecGRC is trained in the same way as GRC, using an entire utterance to obtain a final context $d_{u,T}$ according to Eqs.~\eqref{GRC_context}-\eqref{GRC_recursive}.
At the inference phase, for each decoder step $u$, DecGRC decides an endpoint $t_{end}$ so that only encoded vectors before the endpoint can contribute to the online context vector 
\begin{equation}
\label{DecGRC_context}
c_{u} = d_{u,t_{end}},
\end{equation}
which is used in place of the GRC context vector in Eq.~\eqref{GRC_context}.}
Assume that there exists an endpoint index $t_{end}$ with which $z_{u,t_{end}}$ has a very small value (e.g. less than 0.001). 
Considering that $z_{u,t}<z_{u,t_{end}}$ holds for all $t>t_{end}$, the difference between $d_{u,t_{end}}$ and $d_{u,T}$ is small, as the numerical change $|d_{u,t}-d_{u,t-1}|$ for $t>t_{end}$ induced by the recursion rule in Eq.~\eqref{GRC_recursive} is negligible if $z_{u,t}$ is small enough. Intuitively, intermediate context vectors roughly converge after the endpoint. 

DecGRC can operate as an online attention method if such an endpoint index $t_{end}$ exists at each decoder step 
and the index can be decided by the model. We experimentally 
observed that DecGRC models adequately learn the alignment between encoded vectors and text output units, and the intermediate context nearly converges after the aligned time index at each decoder step. 
\myhl{Nevertheless, the performance of DecGRC can be degraded due to the mismatch between training and inference, especially when the endpoints are decided to be too early.}
Relevant experimental results are given in Sec.~\ref{sec:attention_analysis}

Accordingly, 
with an online encoder, online inference 
can be implemented via a well-trained DecGRC model. We 
describe the online inference technique in Alg.~\ref{alg:example}, where the endpoint index is decided simply by thresholding the update gate values.

\begin{table*}[t]
\caption{Word error rates (WERs) comparison between attention methods on LibriSpeech dataset.}
\label{LibriSpeechWER}
\begin{center}
\begin{small}
\begin{sc}
\begin{tabular}{|l|l|l|c|c|c|rrrr|r|}
\hline
\multirow{3}{*}{\begin{tabular}[c]{@{}c@{}}Exp. \\ ID\end{tabular}} &
\multirow{3}{*}{\begin{tabular}[c]{c}Attention method \end{tabular}} &
\multirow{3}{*}{\begin{tabular}[c]{@{}c@{}}Param. \\ init. \\from\end{tabular}} & \multirow{3}{*}{\begin{tabular}[c]{@{}c@{}}Is \\ attention\\ online?\end{tabular}} & \multirow{3}{*}{\begin{tabular}[c]{@{}c@{}}Is\\encoder\\online? \end{tabular}} & \multirow{3}{*}{\begin{tabular}[c]{@{}c@{}}Can \\infer\\online?\end{tabular}} & \multicolumn{4}{c|}{WER {[}\%{]}} \\ \cline{7-10} 
 & \multicolumn{1}{c|}{} &  &  &  &  & \multicolumn{2}{c|}{dev} & \multicolumn{2}{c|}{test} \\ \cline{7-10} 
 & \multicolumn{1}{c|}{} &  &  &  &  & \multicolumn{1}{c|}{clean} & \multicolumn{1}{c|}{other} & \multicolumn{1}{c|}{clean} & \multicolumn{1}{l|}{other}  \\ \hline \hline
E1 & GSA & - & \multirow{2}{*}{No} & \multirow{7}{*}{\begin{tabular}[c]{@{}c@{}}\\\\\\No\\\\(BiLSTM)\end{tabular}} & \multirow{12}{*}{No} & \multicolumn{1}{r}{\textbf{4.77}} & \multicolumn{1}{r}{14.11} & \multicolumn{1}{r}{4.92} & 15.15 \\ \cline{1-3} 
E2 & GRC & - &  &  &  & \multicolumn{1}{r}{4.84} & \multicolumn{1}{r}{\textbf{14.06}} & \multicolumn{1}{r}{\textbf{4.88}} & \textbf{14.59} \\ \cline{1-4} \cline{7-10} 
E3 & Windowed att. (w=11) & E1 & \multirow{6}{*}{Yes} &  &  & 12.50 & 23.79 & 15.27 & 25.81 \\ \cline{1-3}
E4 & Windowed att. (w=20) & E1 &  &  &  & 5.78 & 14.82 & 5.71 & 15.90 \\ \cline{1-3}
E5 & MoChA (w=2) & - &  &  &  & 6.49 & 17.11 & 6.17 & 18.18 \\ \cline{1-3}
E6 & MoChA (w=8) & - &  &  &  & \textbf{4.74} & 14.20 & 4.95 & 15.32  \\ \cline{1-3}
E7 & \multirow{2}{*}{DecGRC ($\nu$=0.01)} & - &  &  &  & 4.91 & 14.85 & 5.10 & 15.85 \\ \cline{1-1} \cline{3-3}
E8 &  & E2 &  &  &  & 4.97 & \textbf{14.02} & \textbf{4.83} & \textbf{14.90} \\ \cline{1-5} \cline{7-10}   \addlinespace[1pt] \cline{1-5} \cline{7-10}  
E9 & \multirow{2}{*}{GSA } & - & \multirow{4}{*}{No} & \multirow{9}{*}{\begin{tabular}[c]{@{}c@{}}\\\\\\\\Yes\\\\(LC-\\BiLSTM)\end{tabular}} &  & 5.54 & 15.49 & 5.51 & 16.91 \\ \cline{1-1} \cline{3-3}
E10 &  & E1 &  &  &  & \textbf{5.28} & 15.44 & \textbf{5.17} & 16.40 \\ \cline{1-3}
E11 & \multirow{2}{*}{GRC} & - &  &  &  & 6.09 & 16.05 & 6.18 & 16.47 \\ \cline{1-1} \cline{3-3}
E12 &  & E2 &  &  &  & 5.48 & \textbf{15.14} & 5.55 & \textbf{15.88} \\ \cline{1-4} \cline{6-10} 
E13 & \multirow{1}{*}{Windowed att. (w=11)} & E10 & \multirow{6}{*}{Yes} &  & \multirow{6}{*}{Yes} & 12.82 & 24.10 & 15.14 & 26.94   \\ \cline{1-3} 
E14 & \multirow{1}{*}{Windowed att. (w=20)} & E10 &  & & & 5.62 & 15.86 & 5.56 & 16.96   \\ \cline{1-3}
E15 & MoChA (w=2) & E5 &  &  &  & 6.48 & 18.35 & 6.55 & 19.33 \\ \cline{1-3} 
E16 & MoChA (w=8) & E6 &  &  &  & \textbf{5.11} & \textbf{15.10} & \textbf{5.15} & 16.45 \\ \cline{1-3}
E17 & \multirow{1}{*}{DecGRC ($\nu$=0.01)} & E8 &  &  &  & 5.77 & 16.24 & 5.87 & 17.04 \\ \cline{1-2} \cline{3-3}
E18 & \multirow{1}{*}{DecGRC ($\nu$=0.08)} & E12 &  &  &  & 5.79 & 15.67 & 6.04 & \textbf{16.34} \\ \hline 
\end{tabular}
\end{sc}
\end{small}
\end{center}
\end{table*}

\subsection{Computational efficiency of proposed methods}

GRC or DecGRC increases negligible amount of memory footprint, since only one trainable parameter $b$ in Eq.~\eqref{GRC_energy} is added to the standard GSA-based AED model.
\myhl{The computational amount of an attention method is dominated by the score function calculation, as it requires matrix multiplications.
For example in GSA, a fixed-dimensional matrix-vector product is needed to obtain $\tilde{e}_{u,t}$ in Eq.~\eqref{eq:score_exp1} for each $u$ and $t$, which results in $\Theta(TU)$ floating point operations for processing an utterance. Although the softmax operation in Eq.~\eqref{GSA_weights} and the weighted average operation in Eq.~\eqref{eq:GSA_context} also requires $\Theta(TU)$ operations in total, these are negligible compared to the score function calculation since they do not regard matrix-vector multiplications. As a result, the total computational complexity of GSA is $\Theta(TU)$.}

\myhl{Similarly, both GRC and DecGRC requires the score function calculation in Eq.~\eqref{GRC_energy}, having computational complexity of $\mathrm{O}(TU)$. 
}
However, in practice, a speech sequence is linearly aligned with the text sequence on average. As Alg.~\ref{alg:example} only regards to encoded vectors before endpoint indices, the total number of steps in the for loop is typically slightly larger than $TU/2$, if the threshold $\nu$ is set to an appropriate value. Therefore, DecGRC is computationally more efficient than the global attentions such as GRC and GSA \myhl{at the inference phase}. 
The recursive updating in Eq.~\eqref{GRC_recursive} induces negligible amount of computation compared to the whole training or inference process. 
There still exists a room for faster computation by enabling parallel 
computation in time. 
The 
parallel computation can be implemented by utilizing Eq.~\eqref{eq:GSA_context} where $\alpha_{u,t}$ is replaced with $\bar{\alpha}(\mathbf{z}_u)$ in 
Eq.~\eqref{GRC_GSA_dual_function}, instead of Eqs.~\eqref{GRC_context}-
\eqref{GRC_recursive}.
\myhl{
Note that GRC and DecGRC are not the best choices among attention methods in terms of computational complexity. Among the global attention methods, the linearized attention~\cite{katharopoulos2020transformers} features a very low computational complexity of $\Theta(T+U)$ when be used as encoder-decoder attention, which is much smaller than $\Theta(TU)$ of GRC. The computational complexity of an online attention method MoChA~\cite{chiu2018monotonic} is $\Theta(wU)$ where $w$ is the window-size, which is typically far less than $\mathrm{O}(TU)$ of DecGRC. 
Notwithstanding, the encoder-decoder attention's computational amount is minor to the other layers in the encoder and the decoder.
}

The most important fact is that both proposed methods introduce no hyperparameter at the training phase. Thus the proposed methods do not need to repeat training to find a proper value of such a hyperparameter. Though the DecGRC inference in Alg.~\ref{alg:example} introduces a new hyperparameter (i.e., threshold $\nu$) at test phase, the threshold searching on development sets does not take a long time, because the size of the development sets are minor compared to the training set. Hence the total time spent to prepare an ASR system can be saved. Furthermore, the tradeoff between latency and performance can be adjusted by resetting the threshold value $\nu$ at inference phase, unlike the conventional online attention methods~\cite{jaitly2015neural,merboldt2019analysis,chiu2018monotonic}. In these existing methods, the inference algorithms' decision rules on the attention endpoints are determined at the training phase, and 
remains unchanged at the test stage. The experiments on DecGRC with 
different thresholds are demonstrated in Sec.~\ref{sec:ablation_study}.

\section{Experiments}
\label{sec:experiments}

\subsection{Configurations}
\label{sec:configuration}
All experiments were conducted on LibriSpeech dataset
\footnote{The LibriSpeech dataset can be downloaded from \url{http://www.openslr.org/12}.}, 
which contains 16 kHz read English speech with transcription. The dataset consists of 960 hours of a training set from 2,338 speakers, 10.8 hours of a dev set from 80 speakers, and 10.4 hours of a test set from 66 speakers
, with no overlapping speakers between different sets.
Both dev and test sets are split in half into clean and other subsets, depending upon the 
ASR difficulty of each speaker. 
We randomly 
chose 1,500 utterances from dev set as a validation set.

All experiments
\footnote{The scripts for all experiments are available at \url{https://github.com/GRC-anonymous/GRC}.}
shared the same network architecture and training scheme of a recipe 
of RETURNN toolkit~\cite{zeyer2018returnn,zeyer2018comprehensive}, except the attention methods
. Input features were 40-dimensional mel-frequency cepstral coefficients (MFCCs) extracted with Hanning window of 25 ms length and 10 ms hop size, followed by global mean-variance normalization. Output text units were 10,025 byte-pair encoding (BPE) units extracted from transcription of LibriSpeech training set. The $\mathrm{Encoder}(\cdot)$ consisted of 6 BiLSTM layers of 1,024 units for each direction, and max-pooling layers of stride 3 and 2 were applied after the first two BiLSTM layers respectively. For the online $\mathrm{Encoder}(\cdot)$, 6 LC-BiLSTM layers were employed in place of the BiLSTM layers, where 
the future context sizes were 
set to 36, 12, 6, 6, 6, and 6 for each layer from bottom to top and the chunk sizes were 
same as the future context sizes. Both $\mathrm{Score(\cdot)}$ and $\mathrm{MonotonicScore(\cdot)}$ functions were implemented using 
\myhl{the formulation in Eq.~\eqref{eq:score_exp1}} 
and 1,024-dimensional attention key. 
$\mathrm{RecurrentState}(\cdot)$ was implemented with an unidirectional LSTM layer with 1,000 units.  $\mathrm{ReadOut(\cdot)}$ consisted of a max-out layer with 2$\times$500 units, followed by a softmax output layer with 10,025 units. 
\myhl{Every model contains a total of 188~M parameters both for BiLSTM and LC-BiLSTM encoder architecture, except that every MoChA-based model has 191~M parameters.}
 
Weight parameters were initialized with Glorot uniform method~\cite{glorot2010understanding}, and biases were initially set to zero. 
Optimization techniques were utilized during the training: teacher forcing, Adam optimizer, learning rate scheduling, curriculum learning, and the layer-wise pre-training scheme. 
Briefly, the models were trained for 13.5 epochs using a learning rate of 8$\times$10$^{-4}$ with a linear warm-up starting from 3$\times$10$^{-4}$ and the Newbob decay rule~\cite{zeyer2017comprehensive}. Only the first two layers of the $\mathrm{Encoder(\cdot)}$ with half-width (i.e., 512 units for each direction) were used at the beginning of training\myhl{. Then once every 0.25 epoch from 0.75 epoch until 1.5 epoch, a new layer was inserted on the top of the encoder and $1/8$ original width (i.e., 128 units for each direction) of new units are added to each layer.
Finally, the width and the number of layers increased to the original size at 1.5 
epoch.} 
The CTC multi-task learning~\cite{kim2017joint} with a lambda of 0.5 was employed to stabilize the learning, where CTC loss is measured with another 10,025-units softmax layer on the top of $\mathrm{Encoder(\cdot)}$.
For the models which began the learning from parameters of a pre-trained model, the layer-wise pre-training was skipped. Every model was regularized by applying dropout rate 0.3 to $\mathrm{Encoder(\cdot)}$ layers and the softmax layer and employing label smoothing of 0.1. For each epoch of the training, both cross-entropy (CE) losses and output error rates were measured 20 times on the validation set with teacher forcing. 
During the inference phase, model with the lowest WER on the dev-other set among all checkpoints was selected as the final model, and performed beam search once on the dev and test sets with a beam size of 12.

We trained MoChA models for 17.5 epochs with five times longer layer-wise pre-training to make them converge. A small learning rate of 1e-5 was used for training windowed attention models as in~\cite{merboldt2019analysis}. Though the numbers of total epochs for different experiments were not the same, each model was optimized to 
converge and showed negligible improvements after that.

\subsection{Performance comparison between attentions}
All experimental results are summarized in Tbl.~\ref{LibriSpeechWER}. For each experiment, we performed two trials of training with the same configuration and chose a model with the lowest word-error-rate (WER), a word-level Levenshtein distance divided by the number of ground-truth words, on dev-other set. 

In E1 to E2 and E9 to E12, GRC showed better performance than the other attention methods on test-other set, showing 3.7\% and 3.2\% relative error-reduction rate (RERR) compared to GSA when evaluated on BiLSTM and LC-BiLSTM encoder, respectively. 

In E3 to E6 and E13 to E16, performances of the conventional online attentions, i.e., windowed attention and MoChA, 
were shown to be highly dependent on a choice of window size hyperparameter $w$. On the other hand, DecGRC is trained without any additional hyperparameter and only involves a threshold $\nu$ at the inference phase.

In E3 to E8 and E13 to E18, DecGRC outperformed the conventional online attention techniques on BiLSTM encoder. With LC-BiLSTM encoder, the performance of DecGRC on test-other set surpassed the conventional attentions including GSA, while the scores on test-clean set were worse than the competitors. 
The overall performance of GRC and DecGRC is degraded on LC-BiLSTM compared to their preferable performance on BiLSTM, which was conjectured to be caused by the following aspect of the proposed methods; 
$\boldsymbol{\bar{\alpha}}({\mathbf{z}_u})$ 
in 
Eq.~\eqref{GRC_GSA_dual_function} has a dependency on update-gate values of the future time-steps. Therefore using a short future receptive field of LC-BiLSTM may affected the degradation.

\definecolor{blindr}{rgb}{0.832,0.367,0.} 
\definecolor{blindg}{rgb}{0.,0.617,0.449} 
\definecolor{blindb}{rgb}{0.,0.445,0.694} 
\definecolor{blindy}{rgb}{0.797,0.473,0.652} 
\subsection{Optimization speed}

\begin{figure}[t]
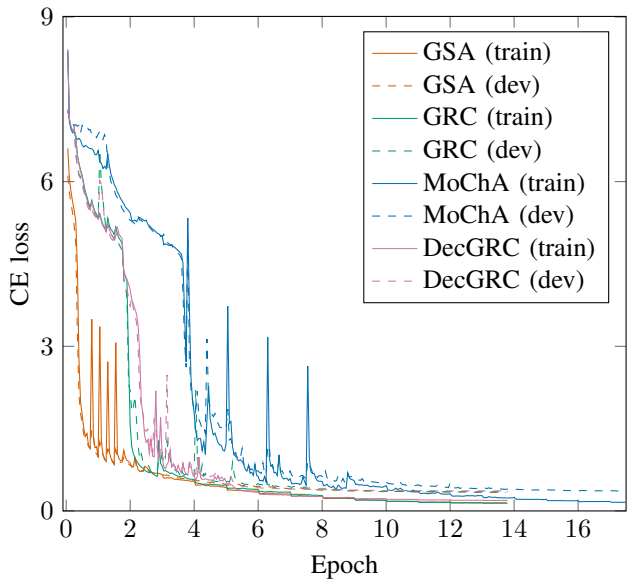

\begin{center}

\caption{Cross-entropy loss curves of various attention methods. All the models were trained from scratch (w/ BiLSTM encoder).}
\label{train_dev_loss}
\end{center}
\end{figure}

The cross-entropy loss curves on training and dev set in E1, E2, E6, and E7 are depicted in Fig.~\ref{train_dev_loss}. The model based on each attention method was trained from scratch until convergence, with a few spikes in its training loss curve. 
\myhl{These spikes in the loss curve are caused by the layer-wise pre-training algorithm described in Sec.~\ref{sec:configuration}. Every time a new layer and units are inserted to the encoder, the training loss temporarily shows rapid increase, because the newly inserted network parameters are not trained yet.}

\myhl{Overall, GRC and DecGRC showed faster from-scratch training speed than MoChA, but slower than GSA.} 
DecGRC converged slightly later than GRC. MoChA showed the slowest optimization speed, which was partly due to the 5 times longer layer-wise pre-training scheduling than the others. Such long pre-training was employed to stabilize the training of MoChA, whereas the both GRC and DecGRC
successfully converged with the standard pre-training.
Note that the longer pre-training of MoCh\myhl{A} was adopted because it 
had failed to converge with a short pre-training in our initial experiments.
The relatively stable learning of the proposed methods over MoChA can be explained in relation to sMoChA, as described in Sec.~\ref{relation_sMoChA}; 
the sMoChA stabilized the training of MoChA by utilizing a modified selection probability formula, which is actually almost similar to the attention weight $\boldsymbol{\bar{\alpha}}({\mathbf{z}_u})$ of GRC in 
Eq.~\eqref{GRC_GSA_dual_function}.

\subsection{Attention analysis}
\label{sec:attention_analysis}
\begin{figure}[t]
\centerline{\includegraphics[width=9.25cm]{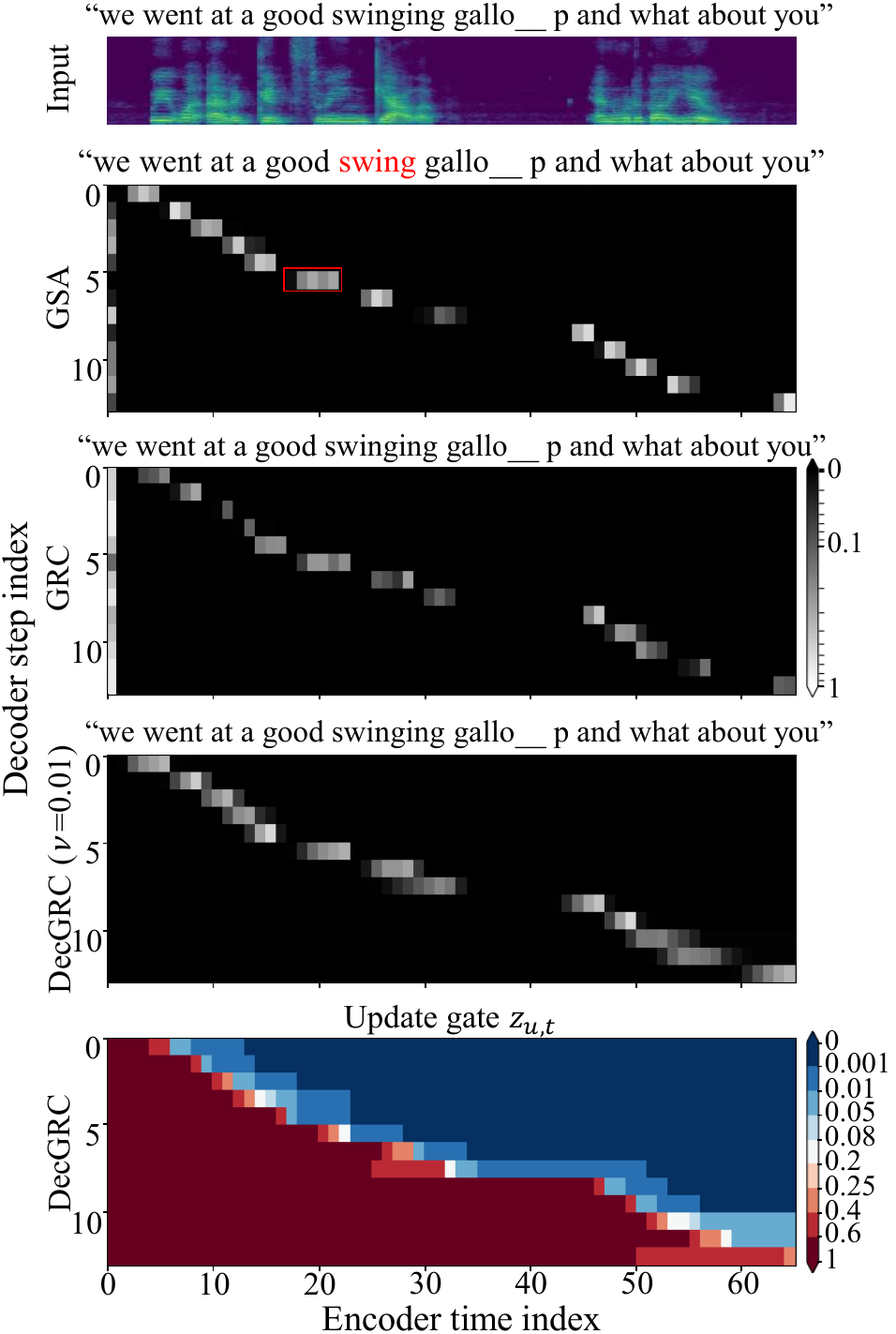}}
\caption{An input spectrogram, attention plots with the output BPE sequence of GSA (E1), GRC (E2), and DecGRC (E8), and the update gates of the DecGRC, from top to bottom. All results were obtained with BiLSTM encoder on
an utterance 8254-84205-0009 in dev-other set. The update gates were obtained with teacher forcing, and the attention plots were results of the beam search w/ beam size 12. ``\_\_" was inserted after a BPE unit end if it was not a word-end. } 

\label{attention_plot}
\end{figure}

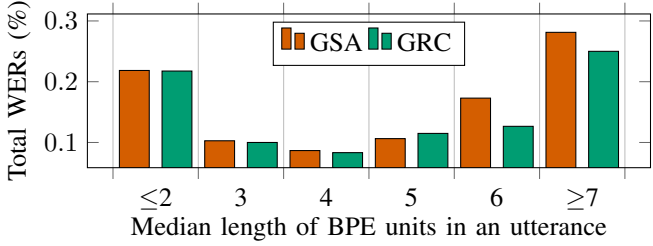
\begin{figure}[t]
\begin{center}
\begin{tikzpicture}
\begin{axis}[
    width=0.5\textwidth,
    height=0.2\textwidth,
	x tick label style={
		/pgf/number format/1000 sep=},
	xlabel=Median length of BPE units in an utterance,
	ylabel=Total WERs (\%),
    x label style={at={(axis description cs:0.5,-0.)},anchor=north},
    y label style={at={(axis description cs:0.08,.5)},rotate=0,anchor=south},
	enlargelimits=0.05,
	legend style={at={(0.5,0.95)},
	anchor=north,legend columns=0},
	ybar interval=0.7,
	xmin=2,
	xmax=8,
	ymin=0.07,
	ymax=0.30,
    xtick={2, 3, 4, 5, 6, 7,8},
    xticklabels={$\leq$2, 3, 4, 5, 6, $\geq$7,8},
]
\addplot[fill=blindr] 
	coordinates {(2,0.21849929211892402)(3,0.10279748360326596)(4,0.08669469039740035)
	(5,0.10634648370497427)(6,0.1729957805907173) (7,0.28125) (8,0)};
\addplot[fill=blindg]
	coordinates {(2,0.21755545068428503)(3,0.10001635955323547)(4,0.08326625525444951)
	(5,0.11492281303602059)(6,0.12658227848101267) (7,0.25) (8,0)};
\legend{GSA,GRC}
\end{axis}
\end{tikzpicture}
\end{center}
\caption{WER for each utterance-wise median length of the BPE units (w/ BiLSTM encoder). WERs for GSA (E1) and GRC (E2) were measured on the test set (i.e., both test-clean and test-other).}
\label{median_length_of_BPE_units_to_WER}
\end{figure}

GRC and DecGRC accurately 
learned alignments between encoded representations and output text units
, as illustrated in Fig.~\ref{attention_plot}. 
An interesting characteristic of GRC was observed that it tended to put much weight on the latter time indices of attention weights, compared to GSA. This can be regarded as an innate behavior of GRC, as the attention weight $\boldsymbol{\bar{\alpha}}({\mathbf{z}_u})$ in Thm.~\ref{thm1} is designed to weigh the latter indices 
when the update gates $z_{u,t}$ have similar value over several consecutive time-indices. 
The latter-time-weighing attribute could be especially effective for a long text unit (e.g., a BPE unit ``swinging'' in Fig.~\ref{attention_plot}), as a long BPE unit often ends with a suffix that might be crucial to distinguish words (e.g., ``-ing", ``-n't", or ``-est" in English). A piece of statistical evidence is presented in Fig.~\ref{median_length_of_BPE_units_to_WER}; GRC outperformed GSA when the median length of BPE was larger than or equal to 6, while it showed similar performance for shorter median lengths.
 
Attention weights of DecGRC tended to be much smoother (i.e., focused on longer time) than GRC and GSA. Such smoothness was hypothesized to be caused by the decreasing update gates, 
which made the model trained to be cautious for a sharp descent of update gate values, as it is irreversible in DecGRC. 
In addition, DecGRC did not attend on the first time index, unlike 
GSA and GRC. It is an intrinsic property of DecGRC, as the earliest update gates have values close to 1 and therefore difficult to carry information to later time. As the initial frames of an utterance usually contain helpful information such as background noise, this might cause DecGRC to be degraded compared to the global attentions.
The last two plots in Fig.~\ref{attention_plot} show that the update gate values of DecGRC mostly changed near the attention region. As the update gates rapidly decreased after the attention region, tight attention endpoints could easily be found by setting the threshold value approximately in a range of [0.001, 0.2]. For instance, with an inference threshold $\nu=0.01$ in Fig.~\ref{attention_plot}, the total number of steps in the for loop in Alg.~\ref{alg:example} was 459, which was approximately 54\% of $TU=13\times65=845$. It implies that insignificant 
time indices were properly ignored during the inference.

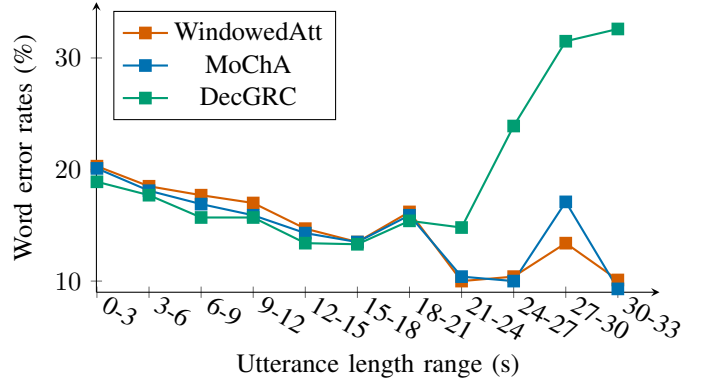
\begin{figure}[t]
\begin{tikzpicture}
\begin{axis}[
width=0.50\textwidth,
height=0.30\textwidth,
axis lines=middle,
xmin=0.99,
xmax=11.8,
ymin=9,
ymax=35,
    x label style={at={(axis description cs:0.5,-0.18)},anchor=north},
    y label style={at={(axis description cs:-0.09,.5)},rotate=90,anchor=south},
    xlabel=Utterance length range (s),
  ylabel=Word error rates (\%),
  xticklabel style = {rotate=-30,anchor=west},
   enlargelimits = false,
  xticklabels from table={uttlen_ablation.dat}{Time},
  xtick=data,
  legend pos=north west]
\addplot[blindr,thick,mark=square*] table [y= WindowedAtt,x=X]{uttlen_ablation.dat};
\addlegendentry{WindowedAtt}]
\addplot[blindb,thick,mark=square*] table [y=MoChA,x=X]{uttlen_ablation.dat};
\addlegendentry{MoChA}
\addplot[blindg,thick,mark=square*] table [y=DecGRC,x=X]{uttlen_ablation.dat};
\addlegendentry{DecGRC}
\end{axis}
\end{tikzpicture}
\caption{Word error rates (WERs) of windowed attention (E14), MoChA (E16), and DecGRC (E18) online models on LibriSpeech test-other dataset for various ranges of utterance lengths, evaluated with LC-BiLSTM encoder. DecGRC model is evaluated with a threshold value of $0.08$.}
\label{uttlen_ablation}
\end{figure}

\myhl{
In Fig.~\ref{uttlen_ablation}, WERs of online attention models are evaluated for various ranges of utterance lengths with LC-BiLSTM encoder. DecGRC models showed better performance than conventional online attention methods for utterances shorter than 21 seconds, while its performance severely degenerated for utterances longer than 21 seconds. We conjectured the performance degeneration of DecGRC for long utterances is fundamentally due to its formulation. According to the recursion rule in Eq.~\eqref{GRC_recursive}, for each decoder step, DecGRC always starts from the first time-index of encoded vectors and processes through the whole sequence until the endpoint is detected, whereas most conventional online attention methods compute the attention weights within a fixed-size window. This indicates that DecGRC has a larger possibility of producing wrong attention context vector than existing online attentions for long utterance, as observed in Fig.~\ref{uttlen_ablation}. The overall performance of DecGRC was better than the others since the utterances longer than 21 seconds is only about 0.5\% of the LibirSpeech test-other set. Notwithstanding, such a low WER problem of DecGRC on long input sequences need to be fixed for better performance, which we would 
solve in future research.
}


\subsection{Ablation study on DecGRC inference threshold}
\label{sec:ablation_study}
We evaluated 
WERs \myhl{and latencies of the proposed online DecGRC model (E18)} 
for different threshold values, and the results are \myhl{plotted in Fig.~\ref{ablation}.}
\myhl{For the latency measure, we employed average lagging~(AL) metric~\cite{ma2019stacl}, which is frequently used to measure the latency of an online sequence-to-sequence model when ground-truth label of input-output time alignment is not given. The AL of an online ASR model on an utterance is obtained as follows~\cite{ma2019stacl}:} 
\pgfplotstableread{
7.208   6.07
2.562   6.08
2.069   6.01
1.952   5.83
1.938   5.79
1.931   5.78
1.901   6.41
1.883   7.27
1.797   15.53
1.761   40.01
}\ablationtable

\begin{figure}[t]
\begin{tikzpicture}
\begin{axis}[
    xlabel={Average lagging (s)},
    xmode={log},
    ymode={log},
    ylabel={Word error rates (\%)},
    xmin=1.7, xmax=2.6,
    ymin=4, ymax=21,
    xtick={1.75, 2.0, 2.25, 2.5, 3.0, 4, 6, 8}, 
    log ticks with fixed point,
    ytick={5,10,20,40},
    legend pos=north east,
    ymajorgrids=true,
    grid style=dashed,
]
\addplot table {\ablationtable}[color=blue] 
    node[pos=0.318, pin=105:0.001]{} 
    node[pos=0.383, pin=75:0.01]{}
    node[pos=0.403, pin=75:0.05]{}
    node[pos=0.41, pin=-105:0.1]{}
    node[pos=0.44, pin=-135:0.2]{}
    node[pos=0.48, pin=75:0.25]{}
    node[pos=0.71, pin=0:0.4]{}
    ;
    \legend{DecGRC}
\end{axis}   
\end{tikzpicture}
\caption{\myhl{Ablation study about the inference threshold of the proposed online model DecGRC (E18) on LibriSpeech dev-clean dataset. The latency measure (average lagging) and WERs were measured with varying inference threshold $\nu$, which is denoted for each node with blue text.}}
\label{ablation}
\end{figure}
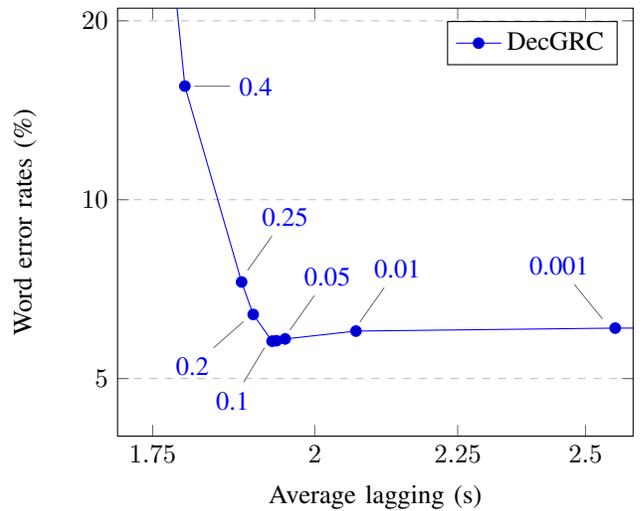
\begin{equation}
\label{average_lagging}
\myhl{ \mathrm{AL}_g (\bold{x},\bold{y}) = \frac{1}{\tau_g(|\bold{x}|)} \sum_{u=1}^{\tau_g(|\bold{x}|)} \Big{\{} {g(u) - (u-1)\frac{|\bold{x}|}{|\bold{y}|}} \Big{\}} }
\end{equation}
\begin{equation}
\myhl{ \tau_g(|\bold{x}|) = \mathrm{min} \big{\{} u \big{|} g(u) = |\bold{x}| \big{\}} }
\end{equation}
\myhl{, where $\bold{x}$ and $\bold{y}$ are acoustic input sequence and output text sequence respectively, and $g(u)$ is a monotonic non-decreasing function of $u$ that denotes the number
of acoustic input frames processed by the encoder when
deciding the $u$-th target text token.
For intuitive notation, we reported the AL value calculated according to Eq.~\eqref{average_lagging} multiplied by the time unit of acoustic input (i.e., 10 ms) in Fig.~\ref{ablation}.}

\myhl{In Fig.~\ref{ablation}, the tradeoff between latency and WER was observed to be adjustable when the threshold $\nu$ is in the range of $[0.1, 1.0]$.}
Setting the threshold to a value larger than \myhl{0.25} 
was found to be detrimental to the performance
, with larger thresholds giving higher WERs. It means that some encoded vectors in the correct attention region were ignored due to the high threshold, as shown in the last two plots of Fig.~\ref{attention_plot}.
Impressively, the best performance was obtained with $\nu$ between \myhl{0.05 and 0.1}
, not $\nu=0$. 
This may be attributed to the fact that the thresholding 
not only reduced the latency, but also eliminated undesirable updates after the correct attention region. With thresholds higher than the best-performing threshold, the latency could be further reduced by taking the performance penalty, and vice versa.

\myhl{After the training end, a DecGRC model needs extra searching to find a threshold that provides the best tradeoff between latency and performance. Nevertheless, the threshold searching time is insignificant compared to the training time.}
The beam search inference on the dev set took less than 15 minutes using a single GPU, 
the time spent for the tuning process of the threshold was no more than 2.5 hours
, which is 
much shorter than 
the model training time; a single epoch of training took 9 hours on average, and the total time for training a model from scratch was more than 5 days. 

\section{Conclusion}
\label{sec:conclusion}
We proposed a novel softmax-free global attention method called GRC, and its 
variant for online attention, namely DecGRC. 
Unlike the conventional online attentions
, DecGRC introduces no additional hyperparameter to be tuned at the training phase. 
Thus DecGRC does not require multiple trials of training, 
saving time for 
model preparation. Moreover at the inference of DecGRC, the tradeoff between ASR latency and performance can be controlled by adapting the scalar threshold which 
is related to the attention endpoint decision, whereas the conventional online attentions are not capable of 
changing the endpoint decision rule at test phase.
Both GRC and DecGRC showed comparable ASR performance to the conventional global attentions.


For further research, the proposed attention methods will be investigated in various applications which leverage AED models. We are particularly interested in applying DecGRC to simultaneous machine translation~\cite{arivazhagan2019monotonic} and real-time scene text recognition~\cite{liu2018squeezedtext}, where the latency can be reduced by exploiting 
an online attention method.

\ifCLASSOPTIONcaptionsoff
  \newpage
\fi

\bibliography{IEEEexample}
\bibliographystyle{IEEEtran.bst}







\end{document}